\documentclass[12pt]{iopart}

\usepackage{graphicx}
\usepackage{hyperref}
\usepackage{lipsum}
\usepackage{multirow}

\begin{document}

\title[Manually driven harmonic oscillator]{Manually driven harmonic oscillator}

\author{M N S Silva and J T Carvalho-Neto}

\address{Departamento de Ci\^{e}ncias da Natureza, Matem\'{a}tica e Educa\c{c}\~{a}o, Universidade Federal de S\~{a}o Carlos, Caixa Postal 153, 13600-970, Araras, SP, Brasil}
\ead{jteles@ufscar.br}
\vspace{10pt}
\begin{indented}
\item[]November 2019
\end{indented}

\begin{abstract}

Oscillations and resonance are essential topics in physics that can be explored theoretically and experimentally in the classroom or teaching laboratory environments. However, one of the main challenges concerning the experimental study of resonance phenomena via forced oscillations is the control of the oscillation frequency, which demands an electronic circuit or a fine tuned coupled mechanical system. In this work, we demonstrate that, in what concerns the physics teaching, such demanding accessories are not necessary. The forced oscillations can be implemented by the teacher's hand guided by an oscillating circle displayed in a web application loaded in a smartphone. The oscillations are applied to an ordinary spiral toy. Qualitative, as well quantitative, proposals are explored in this work with excellent results.

\end{abstract}

\ioptwocol

\section{Introduction}

Oscillations and waves are essential contents for understanding various natural phenomena ranging from basic physics to applied sciences. Specifically, the harmonic oscillators provide great explanatory power as they have analytical solutions in various damping and excitation regimes \cite{thornton2003}.

In addition to the theoretical description of these systems, didactic and illustrative experiments can be performed in the classroom environment, providing a fairly complete description of oscillatory phenomena \cite{macleod1969}. Regarding the damped and driven harmonic oscillator, the electromagnetic and electromechanical systems are the most versatile to work with and have many technological applications. Consequently, they are greatly explored for the teaching of forced oscillations and resonance \cite{macleod1969,jarvis1975,jones1995}. Nowadays, technologies such as smartphone accelerometers have been widely used to study mechanical oscillators \cite{kuhn2012,carlos2013,tuset2015,gimenez2017}. Optical sensors and counters are also employed, some of which use reasonably complex electronic circuits \cite{toyotomi1994,ng2005,wadhwa2009}.

There are few purely mechanical implementations for the teaching of the driven oscillator. As an example, in \cite{ong1972} the author uses the oscillations of a heavy pendulum to drive oscillations on a lighter one. Usually, electrical motors are coupled to spring-mass or pendulum systems in order to drive the oscillations \cite{boving1983,goncalves2017}. These devices, while ingenious and some easy to construct, require time-consuming experimental preparations that can inhibit widespread use in the classroom.

In this work, our purpose is to demonstrate that the excitation and reading of a purely mechanical driven harmonic oscillator aimed at teaching oscillations and resonance can be fairly well performed with a simple spiral toy \cite{pendrill2005, pendrill2014} driven by the own teacher's hand. Moreover, with just few more add-ons, it is possible to quantitatively explore the oscillator amplitude response as function of the external force frequency in different damping conditions. The only necessary technological resource is a smartphone loaded with our open source oscillator web application, which serves as a reference in swinging the spiral toy. A multimedia projector is useful just in case of performing quantitative analysis for the whole classroom.

\section{\label{sec:materials} Materials}

The materials used in this work consisted of a spiral toy and a web application called \textit{harmetronome} \cite{harmetronome2019} that we developed specifically for this work (the name makes reference to the metronome that helps musicians to keep the pace at a chosen beat rate). The web application consists of a circle that executes simple harmonic oscillation. The oscillation frequency or period can be set to a specific value or can be swiped in a predefined range.

The spiral toy consisted of a plastic helical spring with 7.6~cm in external diameter and weighting 0.84~g per loop. Optionally, metal nuts and plastic discs can be attached to the end of the spiral toy in order to change its mass and damping factor, respectively.

%The spiral toy consisted of a 7.5 loop plastic helical spring with 6.5~cm in diameter and weighting 5.90~g. Optionally, metal nuts and plastic discs can be attached to the end of the spiral toy in order to change its mass and damping factor, respectively.

\section{Applications}

\subsection{\label{ssec:qualitative} Qualitative demonstration of resonance phenomenon}

\begin{figure}
	\caption{\label{fig_1} Snapshots of the forced oscillations on the spiral toy (a) below resonance at $\omega=2.6$~rad/s, (b) on resonance at $\omega=4.6$~rad/s, and (c) above resonance at $\omega=6.6$~rad/s.}
	\centering
    \includegraphics[width=1.0\columnwidth]{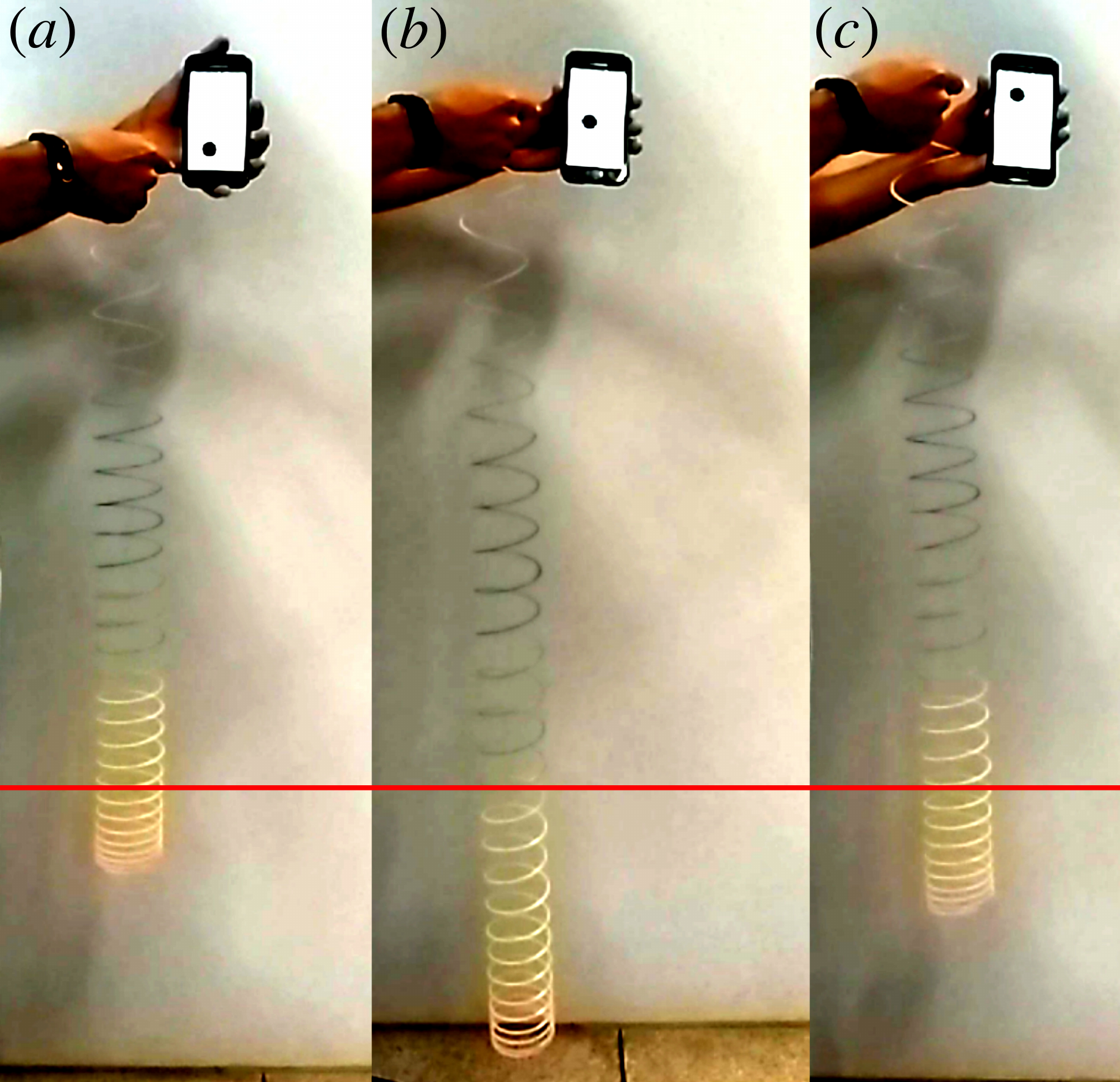}
\end{figure}

This is the simplest propose. The mode of operation is illustrated in figure~\ref{fig_1} \cite{qualitative2019}. The teacher holds the smartphone (loaded with the harmetronome app) with his hand and one end of the spiral toy with the other hand. Both the smartphone display and the spiral toy must be well seen by the students. Then, the hand that holds the spiral toy tries to keep in pace with the oscillating circle being displayed in the smartphone. With such scheme we want to implement a driven harmonic oscillator in order to explore the qualitative aspects related to harmonic oscillations and resonance. For that, we propose the following didactic sequence:

\begin{enumerate}
    \item The teacher presents the concepts associated with the natural frequency of oscillation of a system and resonance phenomenon.
    \item In order to give a simple example of such a system, he keeps one end of the spiral toy motionless with his own hand and puts it to oscillate in the vertical position by displacing and releasing the bottom end.
    \item \label{it:measure} The students are asked to count a number of oscillations of the spiral toy while the teacher measures the elapsed time with the aid of the smartphone chronometer. With that measurement, they find the empirical value for the system natural period of oscillation.
    \item In what follows, the teacher chooses a series of periods of oscillation to be displayed in the harmetronome app: some above, some bellow, and one at the just measured natural period. For each case, the teacher applies the forced oscillations on the spiral toy as described in the earlier paragraph.
\end{enumerate}

Using the procedure~\ref{it:measure} described above, we measured a natural frequency of oscillation of 4.6~rad/s. Therefore, we choose the following frequencies for the manually driven oscillations: 2.6~rad/s, 4.6~rad/s, and 6.6~rad/s. Figures \ref{fig_1}(a) to \ref{fig_1}(c) show snapshots took at the moment that the spring bottom end was at its lowest position for each oscillation frequency. The red line was drew to indicate the spring equilibrium position. That is, the spiral toy bottom end oscillates symmetrically below and above the red line. An unloaded spiral toy with 26 and one fourth loops was used in this demonstration.

As can be seen in figure~\ref{fig_1}, the expected behavior is very well reproduced, despite the inherently inaccurate movement performed by the teacher's hand. Bellow and above the natural frequency, the spring oscillation amplitudes are smaller than the amplitude at the natural frequency. Additionally, at $\omega=2.6$~rad/s we can see that the lowest spring position coincides with the harmetronome lowest postion, which corresponds to a zero phase difference between them. On the contrary, at $\omega=6.6$~rad/s, when the spring bottom end is in the lowest position, the harmetronome is in its highest position, which corresponds to a 180$^{\circ}$ phase difference between them. On resonance, at $\omega=4.6$~rad/s, the harmetronome is in its middle position, which corresponds to a 90$^{\circ}$ phase difference. It is important to note that since the damping factor is significantly lower than the natural frequency, the resonance frequency and the natural frequency are very close to each other.

\subsection{\label{ssec:resonance} Determination of the resonance curve}

It is possible to quantitatively study the oscillator response to an external force by running the harmetronome web application on a computer and projecting it on the wall by means of a multimedia projector. Besides the oscillating circle, there is the option to show a vertical scale against which the shadow of a hanging mass is project and its position can be read by the students. Therefore, as the teacher swings the top end of the spiral toy in pace with the harmetronome circle, the students take note of the minimum and maximum position reached by the hanging mass for each frequency of oscillation choose by the teacher. At the end, the amplitude of oscillation for each frequency of the external force is calculated by taking the difference between the maximum and minimum positions.

In figure 2 \cite{quantitative2019} is shown a picture of the teacher executing the experimental demonstration where the harmetronome application is projected on the blackboard. In order to make the experiment realization more comfortable for the teacher, the spiral toy is attached to the top of a toy car, which in turn is moved by the teacher against the blackboard and a straight piece of wood. The piece of wood helps to constrain the spring movement along a straight line.

\begin{figure}
	\caption{\label{fig_2} Quantitative demonstration of the driven harmonic oscillator. With the left hand, the teacher holds a white piece of wood to vertically constrain the movement executed by its right hand. The shadow of the hanging washer can be seen against the projected scale and the oscillating circle to the left of the scale.}
	\centering
    \includegraphics[width=1.0\columnwidth]{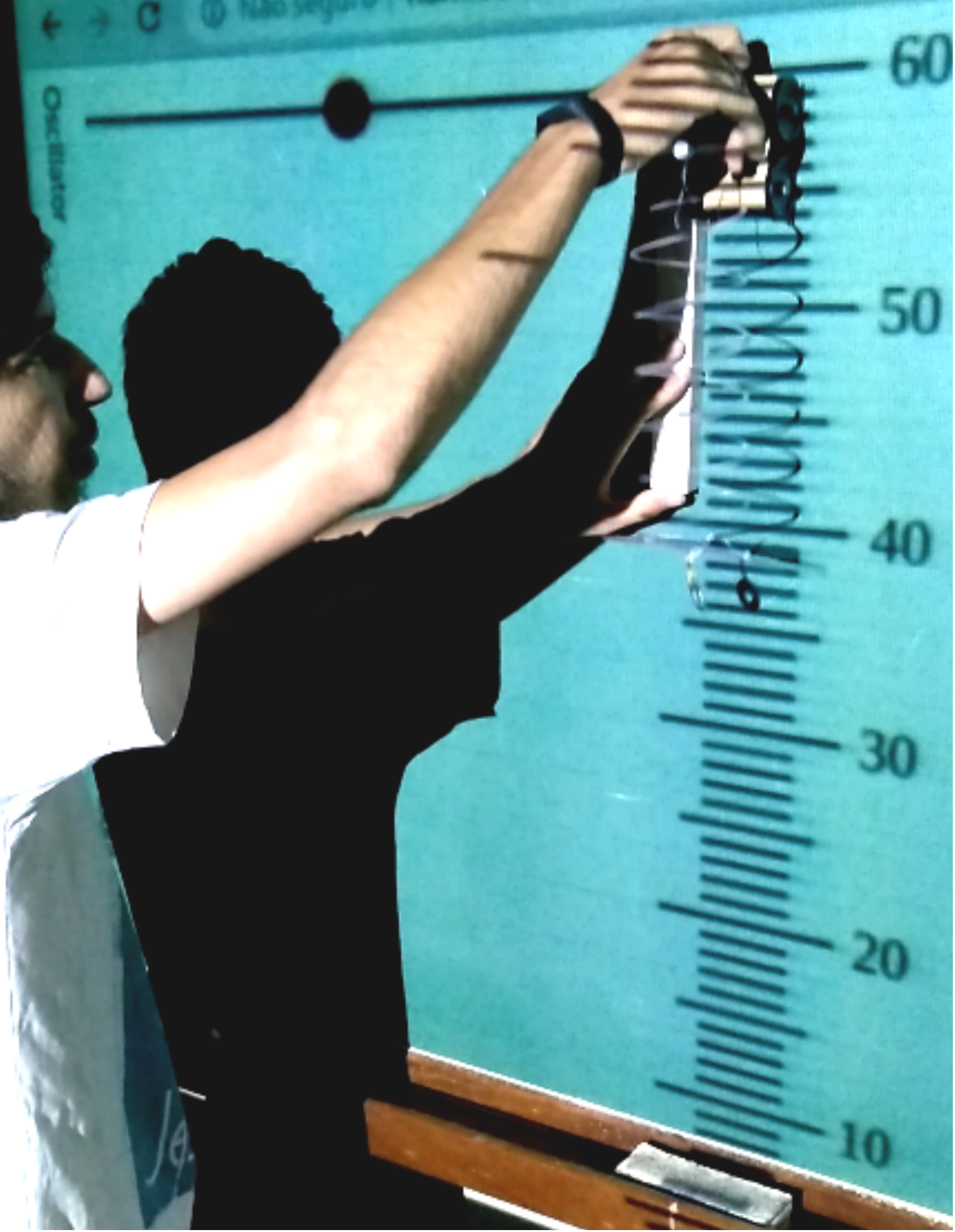}
\end{figure}

With the purpose to demonstrate the effectiveness of this propose, we performed this demonstration to 34 students of a first year class of a Brazilian undergraduate physics teaching course. A piece of six and a half loops of the spiral toy was used. Additionally, a mass of 16.5~g and a 78.5~cm$^2$ plastic disc were attached to the spring end in order to lower the resonance frequency and the maximum oscillation amplitude, respectively. This procedure was important to avoid the spring reaching its minimum elongation. The plastic disc was cut from the cover of a polypropylene folder.

\begin{figure}
	\caption{\label{fig_3}(a) Experimental measurements of the spiral toy resonance demonstration shown in figure~\ref{fig_2}. Each continuous curve was measured by one of the 34 students. (b) The average resonance curve taken from the data of the individual students. The dots correspond to the average values and the error bars to the standard deviations.}
	\centering
    \includegraphics[width=1.0\columnwidth]{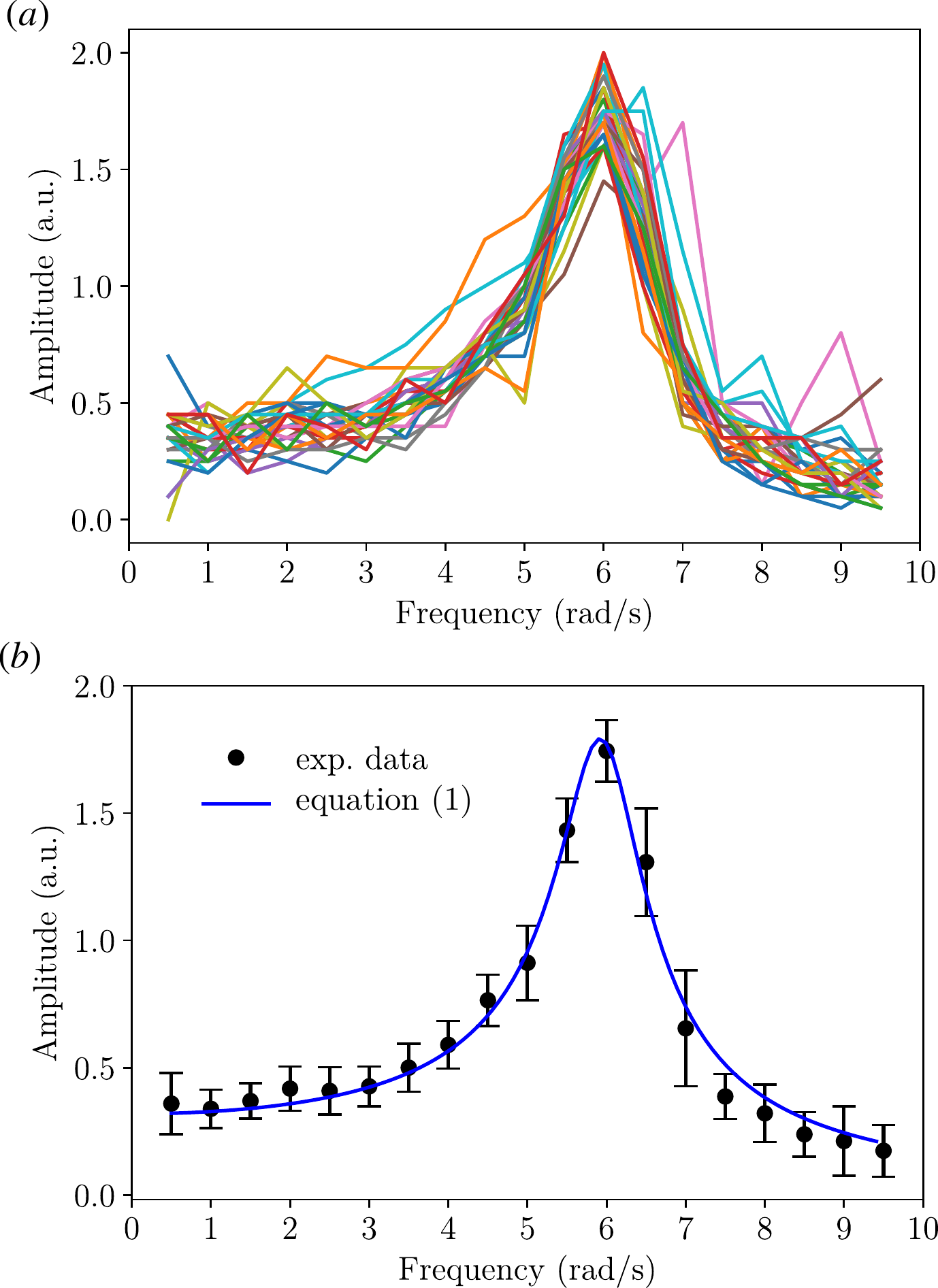}
\end{figure}

Figure~\ref{fig_3}(a) shows the graphic containing the resonance curves recorded by all students. Figure~\ref{fig_3}(b) shows the graphic of the resonance curve obtained by the average of all curves shown in figure~\ref{fig_3}(a). The amplitude of oscillation $A$ of the harmonic oscillator as function of the external force frequency $\omega$ is given by \cite{thornton2003}:
\begin{equation}\label{eq:resonance}
    A(\omega)=\frac{A_0}{\sqrt{\left(\omega_{0}^{2}-\omega^{2}\right)^{2}+4\omega^{2}\beta^{2}}}\;,
\end{equation}
where $A_{0}=F_{0}/m$, $\beta=b/2m$, $\omega_{0}=\sqrt{k/m}$, $F_{0}$ is the external force amplitude, $m$ is the harmonic oscillator mass, and $b$ is the drag coefficient. The continuous curve in figure 3(b) corresponds to the fitting of equation~(\ref{eq:resonance}) to the students measurements.

As result, the following parameters were obtained: $\omega_{0}=(5.97\pm0.03)$~rad/s, $\beta=(0.53\pm0.03)$~rad/s, and $A_{0}=(11.3\pm0.4)$~m/s$^2$. The correlation coefficient of the model to the experimental data was $r=0.991$.

\subsection{Application in the physics lab}

\begin{figure}
	\caption{\label{fig_4} Experiment to be performed in groups of students. The spiral toy is fixed to the top of the guiding white tube. The oscillations are read against the cloth tape in red. The pink plastic ring and the metal washers attached to it change the oscillator damping factor and mass, respectively.}
	\centering
    \includegraphics[width=1.0\columnwidth]{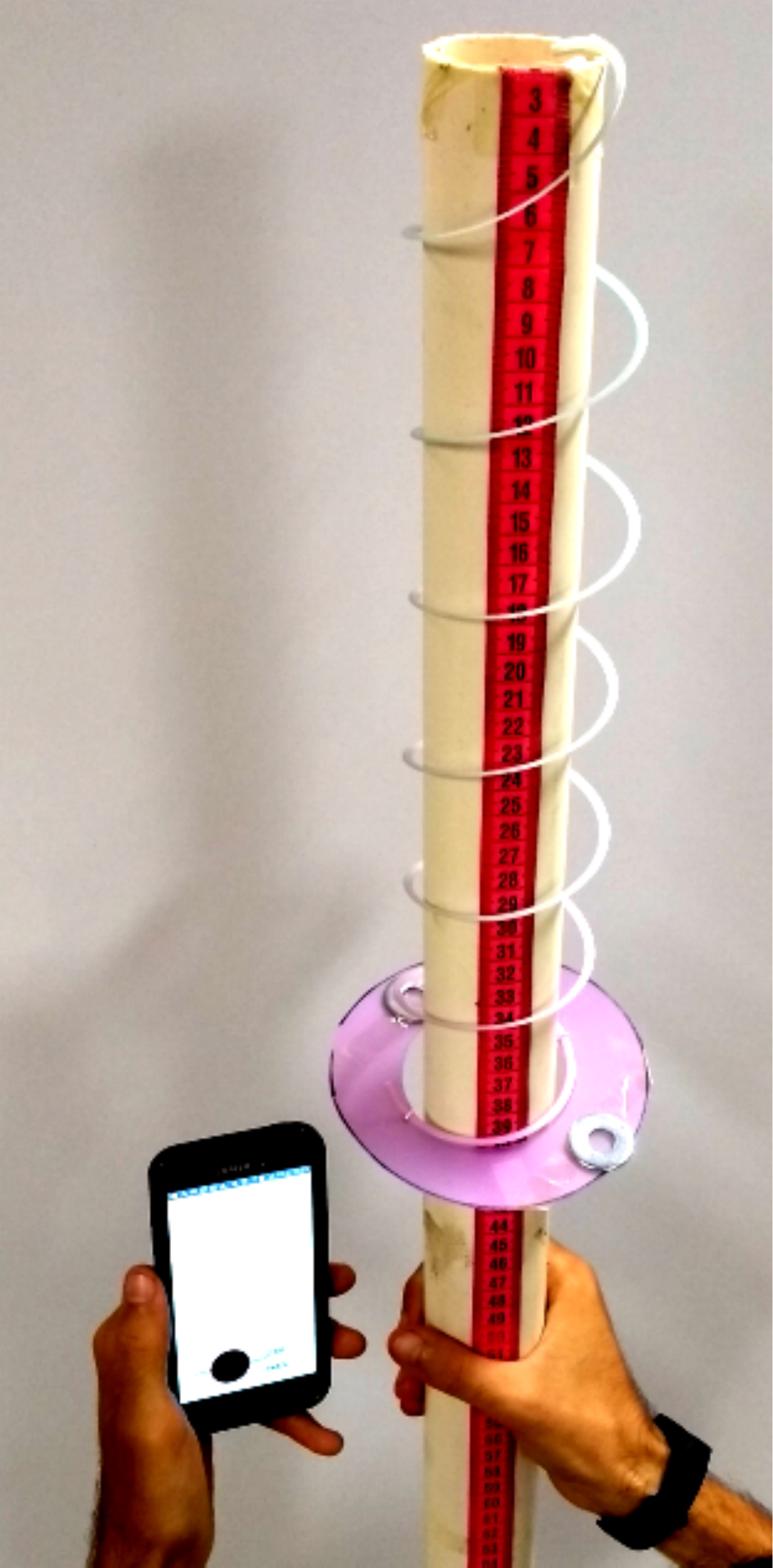}
\end{figure}

Our proposal can also be applied to the physics teaching laboratory, where groups of two or three students can realize the resonance experiment separately. In such case, the spiral toy can oscillate around a rigid tube (e.g. PVC water tube) to which a cloth tape measure is attached. The top end of the spiral toy is fixed to the tube. While one student oscillates the tube in pace with the harmetronome, the other students take note of the minimum and maximum amplitudes reached by the spring as read on the cloth tape. A picture of such configuration is shown in figure~\ref{fig_4}. The purpose of the rigid tube is to constrain the coil movement along a straight line and to hold the scale given by the cloth tape.

\begin{figure*}
	\caption{\label{fig_5} Experimental data obtained with the setup of figure~\ref{fig_4} for different values of effective mass $m_\mathrm{eff}$ and ring area $r_\mathrm{A}$. The continuous curves correspond to the fitting using equation~(\ref{eq:resonance}). The fitted parameters are listed at table~\ref{tab1}.}
	\centering
    \includegraphics[width=2.0\columnwidth]{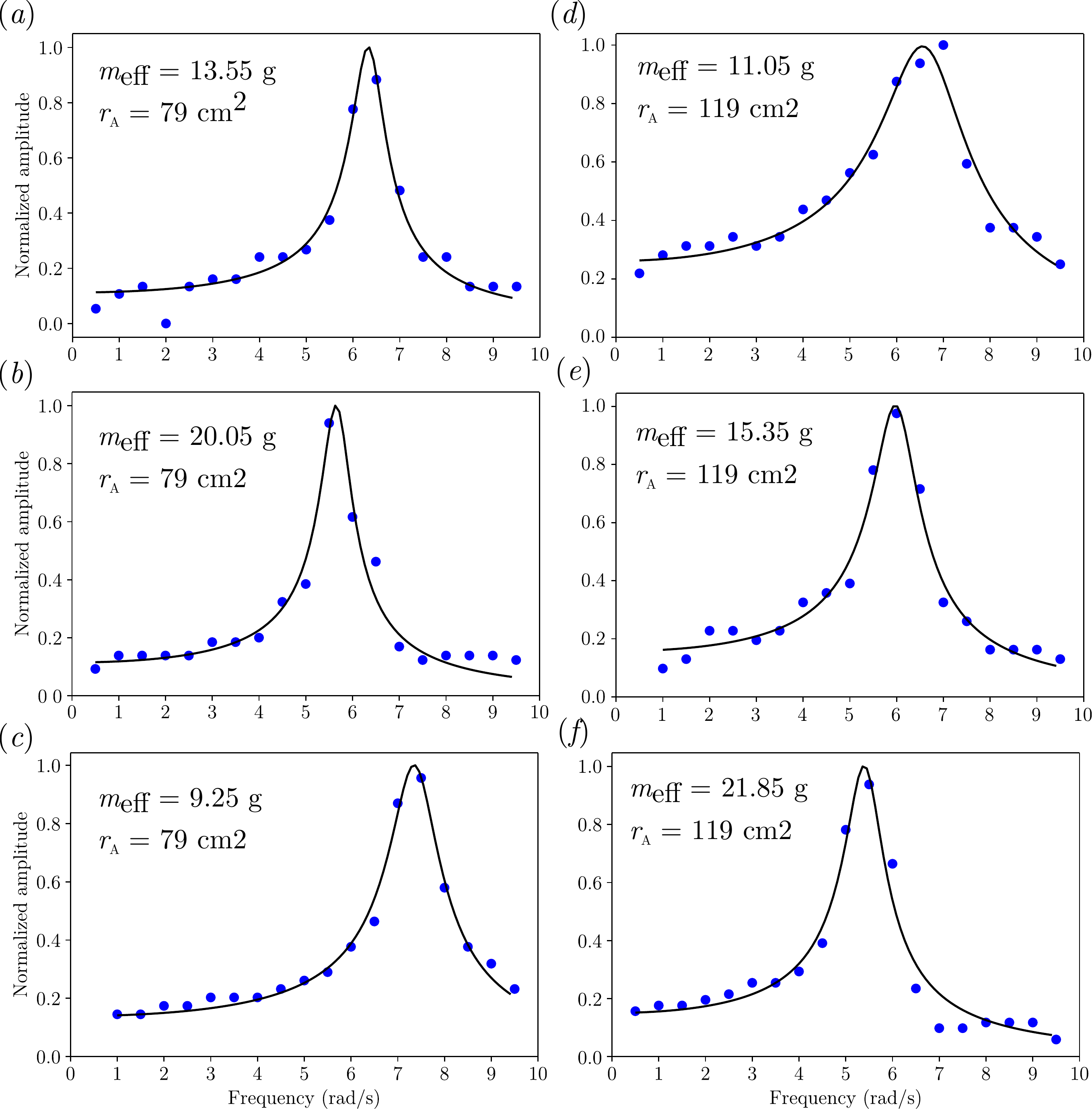}
\end{figure*}

The damping forces and masses can be changed by attaching different plastic rings and metal washers or nuts to the bottom end of the spiral toy. To illustrate the use of this setup, we performed the resonance experiment for the six configurations showed in table~\ref{tab1}. The same piece of six and a half loops spiral toy of section~\ref{ssec:resonance} was used. The resulting resonance curves are shown in figure~\ref{fig_5}(a) to \ref{fig_5}(f). The experimental data is more noisy than the data of figure 3(b) because we did just one measurement for each configuration. However, it could be greatly improved by performing more repetitions and taking their average values.

We also measured the spring elastic constant, $k = 0.57\pm0.01$~N/m, which was obtained using the same apparatus by reading each new spring equilibrium position as function of the attached masses.

In table~\ref{tab1} we compare the undamped natural frequencies $\omega_0$ obtained by the relation $w_{0}=\sqrt{k/m_\mathrm{eff}}$ with this same parameter obtained from the fitting of equation~(\ref{eq:resonance}) to the experimental data in figures 5(a) to 5(f). Since the attached masses, $m$, are in the order of the spiral toy mass ($m_\mathrm{s}=5.9$~g), it was necessary to add a mass correction factor equals to $0.33m_\mathrm{s}$, such that the effective mass was calculated as $m_\mathrm{eff}=m+0.33m_\mathrm{s}$ \cite{fox1970,rodriguez2007}. As can be seen in table~1, the natural frequencies obtained from the two approaches were reasonably close to each other, presenting deviations from 2\% to 7\%. Another correlation that can be observed is between the drag coefficient $b$ and the plastic ring area $r_\mathrm{A}$.

\begin{table*}[h]
\centering
\begin{tabular}{|cccc|cc|}
\hline
\multirow{2}{*}{Figure 5} & \multirow{2}{*}{$m_\mathrm{eff}$ (g)} & \multirow{2}{*}{\begin{tabular}[c]{@{}c@{}}ring area\\ (cm$^2$)\end{tabular}} & \multirow{2}{*}{\begin{tabular}[c]{@{}c@{}}$\omega_{0}=\sqrt{k/m_\mathrm{eff}}$\\ (rad/s)\end{tabular}} & \multicolumn{2}{c|}{From fitting of equation~(\ref{eq:resonance})} \\ \cline{5-6} 
 &  &  &  & \begin{tabular}[c]{@{}c@{}}$\omega_{0}$\\ (rad/s)\end{tabular} & \begin{tabular}[c]{@{}c@{}}$b$\\ (g/s)\end{tabular} \\ \hline
(a) & 13.55 & 79 & 6.50 & 6.34 & 9.7 \\
(b) & 20.05 & 79 & 5.34 & 5.67 & 13 \\
(c) & 9.25 & 79 & 7.86 & 7.38 & 9.5 \\
(d) & 11.05 & 119 & 7.19 & 6.67 & 20 \\
(e) & 15.35 & 119 & 6.10 & 6.01 & 15 \\
(f) & 21.85 & 119 & 5.12 & 5.42 & 18\\
\hline
\end{tabular}
\caption{Driven harmonic oscillator parameters used in the experimental configuration depicted in figure~\ref{fig_4}.}
\label{tab1}
\end{table*}

%All experimental data and numerical analysis presented in this work can be accessed in a public jupyter notebook. The harmetronome code can be accessed in github.

\section{Conclusions}

We have shown that a manually driven harmonic oscillator can be explored in various levels of physics teaching.

In the qualitative approach, the amplitude of oscillation and phase difference behavior as one approaches the system's natural frequency was clearly observed.

A quantitative proposal involving the construction of a resonance curve was applied to a real class of 34 students and the experimental data thus obtained was fitted by the theoretical model with a correlation coefficient of 0.991. One of the main difficulties in this approach is the realization of many different oscillations by a single person (e.g. the teacher) whose arm can get tired. In this case, the teacher may take turns with other students.

At last, an implementation using a rigid tube as a guide to the spiral toy was proposed to be applied to a physics teaching laboratory, where each group of students can make its own quantitative analysis probing the forced damped harmonic oscillator in different conditions.

\section*{References}

\bibliography{references}{}
\bibliographystyle{iopart-num}

\end{document}